\documentclass[sigconf]{acmart}
\usepackage{framed}
\usepackage{color}
\usepackage{algorithm}
\usepackage{algorithmic}
\usepackage{subcaption}
\usepackage{multirow}

\definecolor{shadecolor}{rgb}{0.92,0.92,0.92}
\newcommand{\para}[1]{{\vspace{2pt} \bf \noindent #1 \hspace{1pt}}}

\AtBeginDocument{%
  \providecommand\BibTeX{{%
    \normalfont B\kern-0.5em{\scshape i\kern-0.25em b}\kern-0.8em\TeX}}}

\setcopyright{acmcopyright}
\copyrightyear{2018}
\acmYear{2018}
\acmDOI{10.1145/1122445.1122456}

\acmConference[Woodstock '18]{Woodstock '18: ACM Symposium on Neural
  Gaze Detection}{June 03--05, 2018}{Woodstock, NY}
\acmBooktitle{Woodstock '18: ACM Symposium on Neural Gaze Detection,
  June 03--05, 2018, Woodstock, NY}
\acmPrice{15.00}
\acmISBN{978-1-4503-XXXX-X/18/06}

\begin{document}

\title{GLOW : Global Weighted Self-Attention Network for Web Search}

\author{Xuan Shan}

\affiliation{%
  \institution{Microsoft STCA}
}
\email{xuanshan@microsoft.com}

\author{Chuanjie Liu}
\affiliation{%
  \institution{Microsoft STCA}}
\email{chuanli@microsoft.com}

\author{Yiqian Xia}
\affiliation{%
  \institution{Microsoft STCA}}
\email{yiqxia@microsoft.com}

\author{Qi Chen}
\affiliation{%
  \institution{Microsoft Research Asia}}
\email{cheqi@microsoft.com}

\author{Yusi Zhang}
\affiliation{%
  \institution{Microsoft STCA}}
\email{zhangyusi@pku.edu.cn}

\author{Kaize Ding}
\affiliation{%
  \institution{Arizona State University}}
\email{kding9@asu.edu.cn}

\author{Yaobo Liang}
\affiliation{%
  \institution{Microsoft Research Asia}}
\email{yalia@microsoft.com}

\author{Angen Luo}
\affiliation{%
  \institution{Microsoft STCA}}
\email{anluo@pku.edu.cn}

\author{Yuxiang Luo}
\affiliation{%
  \institution{Microsoft STCA}}
\email{yuxlu@pku.edu.cn}
\renewcommand{\shortauthors}{Shan, et al.}

\begin{abstract}
Deep matching models aim to facilitate search engines retrieving more relevant documents by mapping queries and documents into semantic vectors in the first-stage retrieval.
When leveraging BERT as the deep matching model, the attention score across two words are solely built upon local contextualized word embeddings.
It lacks prior global knowledge to distinguish the importance of different words, which has been proved to play a critical role in information retrieval tasks. 
In addition to this, BERT only performs attention across sub-words tokens which weakens whole word attention representation.
We propose a novel \textbf{Glo}bal \textbf{W}eighted Self-Attention (GLOW) network for web document search. GLOW fuses global corpus statistics into the deep matching model. By adding prior weights into attention generation from global information, like BM25, GLOW successfully learns weighted attention scores jointly with query matrix $Q$ and key matrix $K$.
We also present an efficient whole word weight sharing solution to bring prior whole word knowledge into sub-words level attention. It aids Transformer to learn whole word level attention.
To make our models applicable to complicated web search scenarios, we introduce combined fields representation to accommodate documents with multiple fields even with variable number of instances.
We demonstrate GLOW is more efficient to capture the topical and semantic representation both in queries and documents. Intrinsic evaluation and experiments conducted on public data sets reveal GLOW to be a general framework for document retrieve task. It significantly outperforms BERT and other competitive baselines by a large margin while retaining the same model complexity with BERT. The source code is available at \url{https://github.com/GLOW-deep/GLOW}.

\end{abstract}




\maketitle

\section{Introduction}
\label{sec1}
Nowadays, modern search engines leverage two-stage algorithms to retrieve ideal results from a massive amount of documents in order to obtain milliseconds query response time. The first stage applies a coarse-grained search to quickly select a small set of candidates from billions of documents using low-cost metrics. Then some complex ranking algorithms at the second stage are adopted to prune the results. 
Traditionally, the first-stage retrieval is built on top of an inverted index using keyword match with some query alterations. However, it is hard to cover all the ideal cases and well understand user's intention. If the alteration technique fails to enumerate all the keyword expansions, some ideal documents will be missed. With recent breakthrough in deep learning, web contents can be more meaningfully represented as semantic vectors. Especially for large scale retrieval tasks, vector recall\cite{xiong2020approximate} has been attracting more attention to remedy the disadvantages of traditional keyword-based approach. It leverages high efficient Approximate Nearest Neighbor(ANN)\cite{aumuller2017ann} search algorithms to retrieve relevant results according to the vector distance. Given that the ANN index is supposed to be pre-built ahead of serving, documents have no chance to interact the queries at encoding stage. To achieve this, deep matching models usually adopt a Siamese architecture to embed documents without the help of queries. Traditional examples include DSSM\cite{dssm}, C-DSSM\cite{cdssm} and ARC-I\cite{ARC1}.
Recently, Transformer based models like BERT\cite{devlin-etal-2019-bert} are being widely used as the deep matching model \cite{nogueira2019passage,qiao2019understanding,reimers2019sentence}. However, when leveraging the vanilla BERT, there are three limitations we need to address.  

First, the attention calculation in BERT relies on local context within the single sequence. It fails to capture global information from whole corpus. For example, in the query ``\textit{what are the worst effects of pesticides to nature}", the embedding representation to this query are jointly trained based on query matrix $Q$, key matrix $K$ and value matrix $V$ from entire words without distinction. But it is obvious that ``\textbf{pesticides}" and ``\textbf{worst}" should commit higher attention scores since these two words are more crucial to represent the topic. If we stand at a global statistics view, we do have chances to identity the importance of ``\textbf{pesticides}" and ``\textbf{worst}". These two words rarely appear in other sequences while ``\textbf{what}", ``\textbf{are}" and ``\textbf{effects}" empirically have higher frequencies, one alternative is that we can leverage global statistics features like Inverse Document Frequency(IDF) or BM25\cite{bm25AndBeyond} as signals of global weights to adjust the original attention scores. 

The second issue is that when applying WordPiece embedding\cite{wordpiece}, a natural word may be split into several different tokens, which leads to BERT's attention solely behaving at sub-words level and lacking whole word level interaction. To remedy this limitation, the latest released BERT model has upgraded the mask language model task to whole word level, but it still does not involve weight distinctions across different whole words.   

Thirdly, building a suitable deep matching model to for web document retrieval is challenging, not only because the aforementioned challenges, but also multiple fields of documents should be taken into consideration. There are always multiple sources of textual description (\textit{fields}) corresponding to one web document. Lots of studies \cite{neuralrankingmodelmultiplefields,bm25ForWeightedFields} reveal that different fields contain complementary information. 
Previous studies on deep matching model mainly consider with the single field document or simply concatenate multiple fields as one unified field\cite{ARC1,cdssm,dssm}. Seldom researches propose suitable solutions to adapt multi-fields web document scenarios.
Thus, to obtain a more comprehensive vector representation for the first-stage retrieval, an efficient solution on multi-fields is critical. 


Empirical studies\cite{bm25ForWeightedFields,LabmdaBM25} show global representative features like BM25 well express term importance with global context information. A word with high BM25 score reveals its uniqueness in the corpus. It has been widely adopted in traditional learning to rank tasks, unfortunately seldom studies investigate to integrate it into Transformer as deep matching models. Kim et al. \cite{Kim2019TGSATW} significantly improves speech-enhancement performance by integrating a Gaussian-weight into attention calculation.
Inspired by this, in this paper we introduce GLOW: a Global Weighted Self-Attention network to learn the semantic representations of queries and documents. Specifically, it pre-computes BM25 scores for query terms and document terms respectively, then taking the scores as the global guiding weight when performing self-attention. GLOW leverages a ~30k token vocabulary from WordPiece embedding, while BM25 is usually generated on natural word level, since one natural word may be mapped into different WordPiece tokens, It is vital to pave a way to pass BM25 score from word level to token level. To demonstrate whole word level attention, we propose a whole word weight sharing mechanism to bridge the discrepancy between natural words and WordPiece tokens. Since web documents are described with multiple fields, we further introduce a combined fields representation solution, which successfully differentiates document fields by involving field embeddings. 

To the best of our knowledge, this is the first research work that successfully integrates global statistic information into self-attention based models as a guiding weight.
GLOW significantly improves the search relevance by intrinsic evaluation with Bing's search logs. We also measure GLOW on MS MARCO\cite{bajaj2016ms}, the results and analyses show GLOW is superior in retrieval quality without increasing any model complexity.

To summarize, our contributions are:
\begin{itemize}
    \item We point out that vanilla Transformer may not obtain accurate attention scores in Web search scenario due to lacking global prior knowledge .
    \item We demonstrate a novel deep matching model by integrating global weight information into Transformer and
    the whole word weight sharing successfully bridges the sub-word tokens and full words discrepancies.
    \item We propose the combined fields representation for multi-fields documents. It well handles fields prejudice by field embeddings and consolidates into one vector representation for per document.
    \item We conduct rigorous comparisons over state-of-the-art deep matching models on public dataset and Bing's large scale validation data points. Detailed analysis is also conducted to study why GLOW can achieve better results.
\end{itemize}
\begin{figure*}
\centering
\includegraphics[width=18cm]{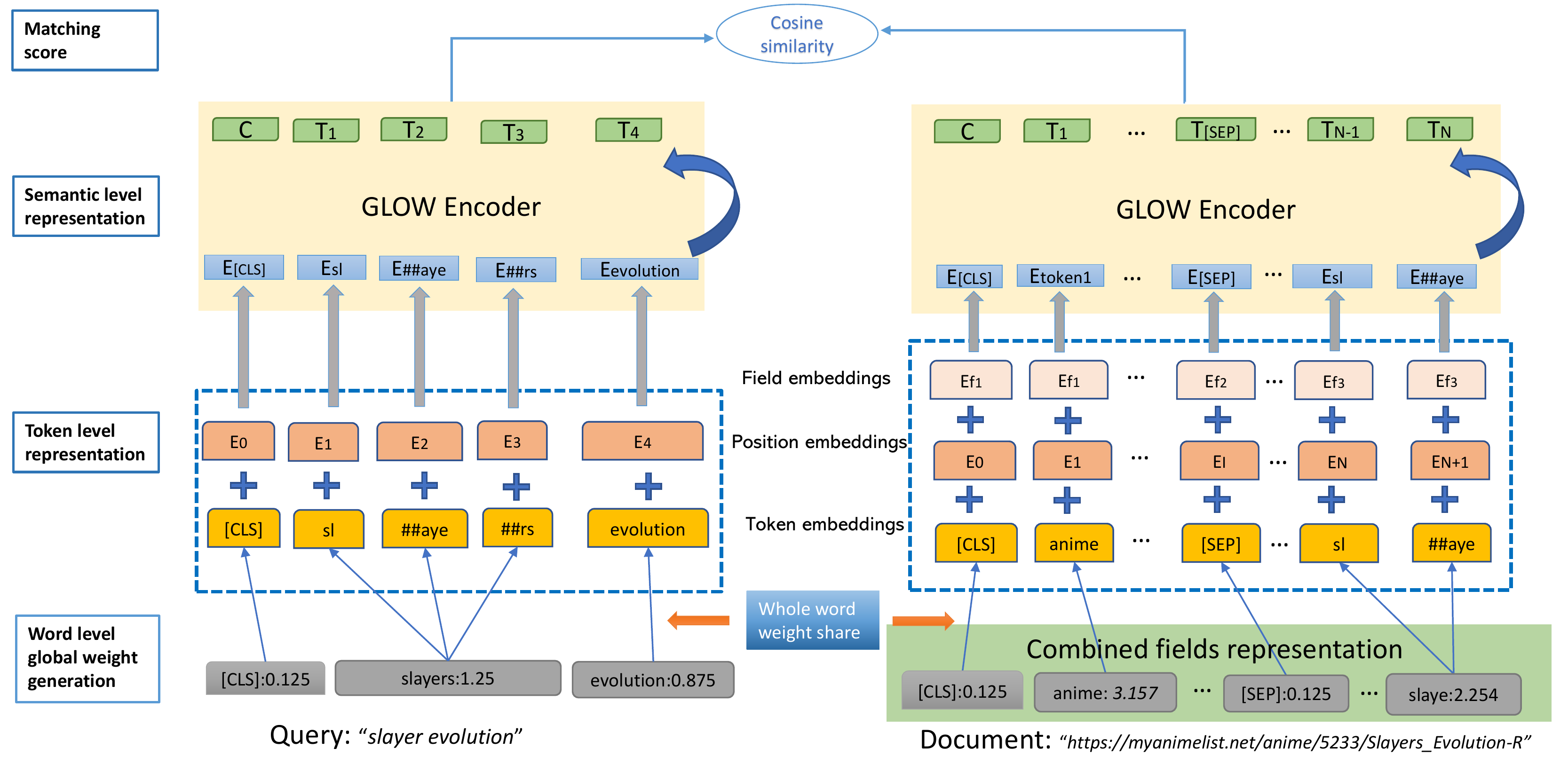}
\caption{Illustration of GLOW. We present a sample query and document to describe the network structure. The core block GLOW Encoder can be repeated many times. The blue arrows between word level weight generation and token level representation indicate the whole word weight sharing methodology. One word might map to single or multiple tokens.}
\label{fig:model_structure}
\end{figure*}

\section{Related work}
\label{sec2}
Recently, a variety of deep matching models have been proposed for the text matching problems, Mitra and Craswell \cite{introToNeuralRanking}gave a detailed introduction about the researches made on information retrieval with deep neural networks. Specifically, Huang et al.\cite{dssm}  developed DSSM with a deep structure that project queries and documents into a common low-dimensional space. Shen et al.\cite{cdssm}  upgraded DSSM to C-DSSM by using the convolution-max pooling operation.
Guo et al.\cite{guo2016deepmatchingmodel} formulated the ad-hoc retrieval as a matching problem and proposed Deep Relevance Matching Model(DRMM). Deep matching models are usually equipped into search engines by a Siamese (symmetric) architecture\cite{cdssm,dssm,ARC1} or an Interaction-focused manner\cite{ARC2,DeepMatch,MatchPyramid}. The major difference between these two architectures lies in when a query interacts with the document, the Siamese approach encodes the query and document separately while Interactive way jointly learns their correlations at the very beginning. For large scale document matching tasks, especially those that depend on vector search, the Siamese approach is preferred since a multitude of documents are supposed to be encoded without the help of queries. To better facilitate document matching tasks, our proposed framework GLOW is built upon Siamese architecture.

In addition to this, pre-train language modeling has been proved to be effective on natural language processing tasks. One of such models, BERT, has been widely applied into retrieval-based tasks like document ranking\cite{yang2019simple} and question answering\cite{yang2019end,nogueira2019passage}.
MS MARCO\cite{bajaj2016ms} is a collection data set for multi-perspective web search tasks. 
The top 10 winners in the leading board all leverage BERT as a basis. Typically, Nogueira et al.\cite{nogueira2019multi} built a multi-stage ranking architecture on BERT by formulating the ranking problem as pointwise and pairwise classification, respectively. Han et al. combined DeepCT retrieval model\cite{dai2019deepcpt} with a TF-Ranking BERT ensemble\cite{tfRanking}. The DeepCT-Index produces term weights that can be stored in an ordinary inverted index for document ranking. 
Observed from another famous information retrieval data set ClueWeb09\cite{callan2009clueweb09}, the announced high results are also trained on Transformer based models. By proposing a generalized autoregressive pretraining method XLNet\cite{Yang2019XLNetGA} claimed its state-of-the-art result, superior to RoBERTa\cite{liu2019roberta} , GPT\cite{radford2018improving} and BERT+DCMN\cite{zhang2019dual}. 
Besides, Yilmaz et al.\cite{yilmaz2019applying} presented Birch, a system that applies BERT to document retrieval via integration with the open-source Anserini information retrieval toolkit to demonstrate end-to-end search over large document collections.
From the document-query matching perspective, Doc2query\cite{doc2query} predicts which queries will be issued for a given document and then expands it with those predictions with a vanilla Transformer model, trained using datasets consisting of pairs of query and relevant documents. ColBERT\cite{khattab2020colbert} introduces a late interaction architecture that independently encodes the query and the document using BERT.
\\Most of these studies consolidate on single field document. Although Zamani et al.\cite{neuralrankingmodelmultiplefields} proposed a deep neural ranking model on multi-fields document ranking. Self-Attention based approaches have not been well studied yet for multi-fields document.





\section{The GLOW FRAMEWORK}
In this section, we first provide the formulation of document retrieval task. Then we introduce a high-level overview of our framework and further describe how we implement each component of the proposed framework. We finally explain how we optimize our deep matching model.
\label{headings}
\subsection{Problem Statement}
The first-stage of document retrieval task can be described as, given one query \textit{q}, the system produces a fixed amount of documents \textit{D} from a mass of candidates. \textit{d} represents one instance from \textit{D}. Since in web search \textit{d} is always equipped with multi fields contents. Let $\mathcal{F}_d = \{F_1, F_2, \cdots, F_k\}$ denote a set of fields associated with the document \textit{d}.

For a single \textit{q}, \textit{d} pair, a deep matching model usually describes a matching score based on the representation of \textit{q} and \textit{d}.
\begin{equation}
match(q,d)=Func(\Phi(q),\Phi(\mathcal{F}_d))
\end{equation}
where $\Phi$ is a model function to map each query and document to a representation vector, and $Func$ is the scoring function based on the similarity between them.
\subsection{Overview of GLOW}
As show in Figure \ref{fig:model_structure}, to fit large scale document matching scenario, GLOW is built on a Siamese manner, we employ two GLOW Encoders to represent queries and documents respectively.
From a horizontal view, GLOW is comprised of four parts, Word level global weight generation, Token level representation, Semantic level representation and Matching score. 
One prerequisite on data preparation before training is that we simply prepend a [CLS] (classification) token to both query and document. For document side, a [SEP] (separating) token is inserted across different fields to constitute the combined fields representation. 

We introduce semantic level representation in Section \ref{sub:weightedattention} and \ref{glow_encoder}, which takes GLOW Encoder by stacking 3 times.
Then we describe how to generate the word level global weight in Section \ref{weight_generation}, what we adopt finally as global weight is BM25. 
The whole word weight sharing is described in Section \ref{wholeWordWeightShare} to explain how we map weight from whole words to WordPiece tokens.
For the token level representation, we use the sum of token embeddings and position embeddings to form the token representation for query side. For document, we introduce the combined fields represention in Section \ref{combinedFiledsRepresentation}, it adds field embeddings to differentiate multi-fields in documents. 
We adopt \textit{cosine similarity} to describe the matching score. Section \ref{optimization} explains how we optimize the matching score with training labels.


\subsection{Global Weighted Self-Attention}
\label{sub:weightedattention}
Let's define $\mathbf{X} \in \mathbb{R}^{d \times T}$ is a $d$-dimensional sequence embedding input of one query or document with length of $T$, $x_{i}$ is the $i$th token in the sequence. $Q$,$K$ and $V$ are matrices initiated by $\mathbf{X}$ multiplying different weight matrices.
The attention score matrix in such a sequence is denoted by $\mathbb{A} \in \mathbb{R}^{T \times T}$. For a token pair $x_{i}$, $x_{j}$, $q_{i}$ and $k_{j}$ are the column selection from $Q$ and $K$ according to $i$ or $j$, its attention score $A_{ij}$ is calculated in Scaled Dot-Product Attention as $A_{ij} = \frac{{q_i}\cdot{k_j^T}}{\sqrt{d}} \label{con:attention}$.
In \cite{AttentionisAllyouNeed}, they claim the attention unit is already a weighted sum of values, where the weight assigned to each value is learned from $q_{i}$ and $k_{j}$.
Whereas, in information retrieval area, it is well equipped with prior knowledge to represent the weight of one word. We enrich the attention calculation with these techniques. Assuming $w_{i} \in \mathbb{R}^{T}$ represents the global weight of the $i$th token in the sequence, $w_{i}$ is a \textbf{non-trainable scalar}. In this paper we use BM25 to represent this global importance. Thus a new weighted attention score $A_{ij}^{w}$ is computed as 
\begin{equation}
\label{weightedSelfAttention}
A_{ij}^{w} = w_{j}\frac{{q_i}\cdot{k_j^{T}}}{\sqrt{d}},  A_{ji}^{w} = w_{i}\frac{{q_j}\cdot{k_i^{T}}}{\sqrt{d}}
\end{equation}
 where $q$ and $k$ share the same shape and only differ in random initialization. Symmetrically, the weighted attention score of $A_{ji}$ can be represented in the right part of Eq. \ref{weightedSelfAttention}. The right part of Figure \ref{fig:model_structure} presents how Weighted Self-Attention works. With importing the weight information and packing all $w_{i}$ into $W$, we formally define \textbf{Global Weighted Self-Attention} (GWSA)  as
\begin{equation}
\begin{aligned}
\begin{split}
    {\rm WeightedSelfAttention}(Q,K,W,V)= \\ {\rm softmax}(W \odot \frac{QK^{T}}{\sqrt{d}})V\label{eq:weightedAttention} 
\end{split}
\end{aligned}
\end{equation}
where $W$ is one dimension vector and its multiplicand is a matrix . $\odot$ represents a Hadamard product by repeating $W$ to perform element-wise multiplication. Eq \ref{sample_hadamard} explains this special operation.
\begin{equation}
\label{sample_hadamard}
W \odot A = (w_i\cdot a_{ij} ) = 
\begin{pmatrix} 
 w_{1} \cdot a_{11} & \cdots & w_{1} \cdot a_{1n} \\
\vdots & \ddots & \vdots \\ 
w_{m} \cdot a_{m1} & \cdots & w_{m}  \cdot a_{mn}
\end{pmatrix}
\end{equation}

\begin{figure}
\includegraphics[width=8cm]{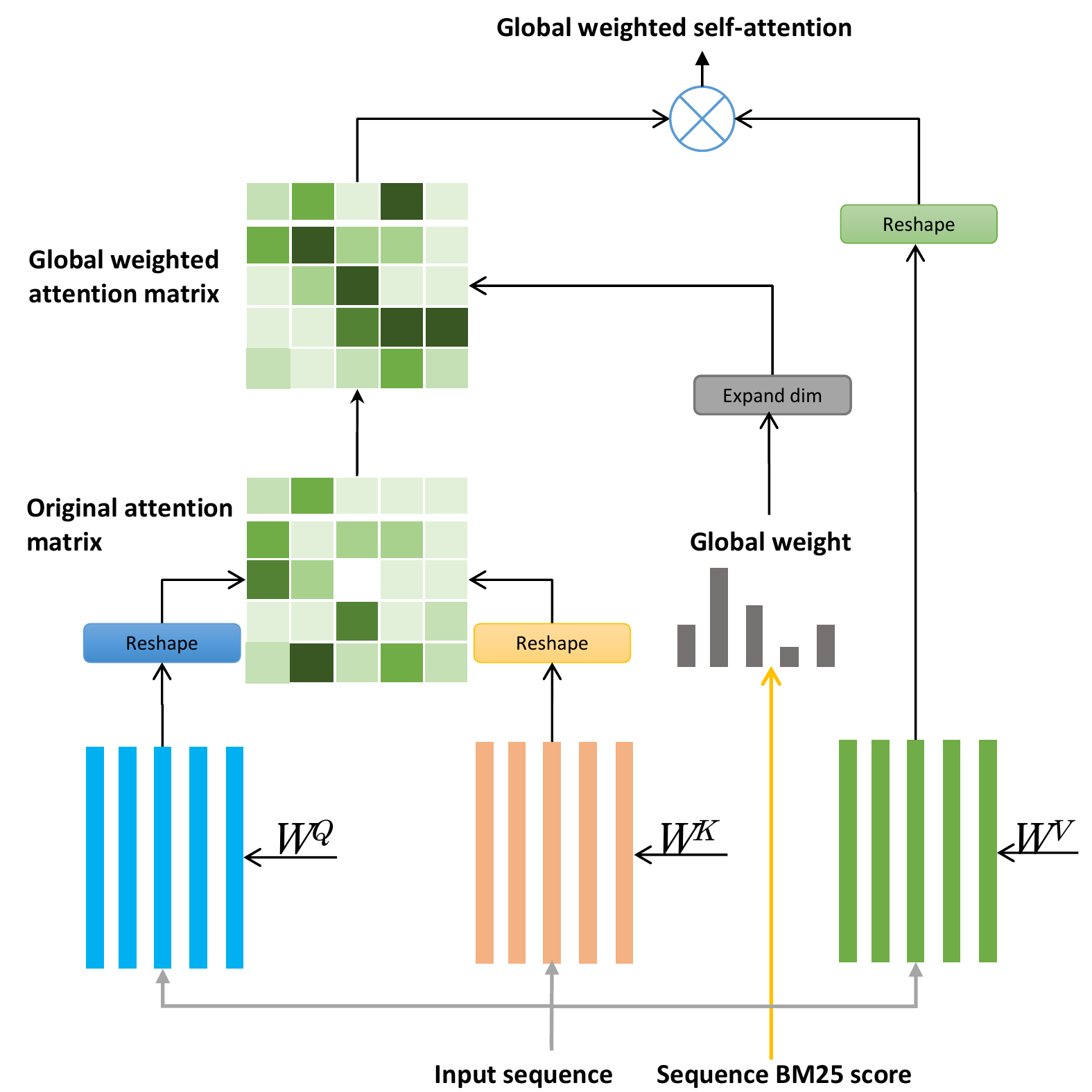}
\caption{Block diagram of the proposed Global Weighted Self-Attention.}
\label{fig:weighted_self_attention}
\end{figure}


\subsection{GLOW Encoder}
\label{glow_encoder}
GLOW Encoder picks this Weighted Self-Attention as its block unit. It is also built upon multi-head structure by concatenating several Weighted Self-Attention instances. With re-scaling by $W^o$, we can get a Complex Weighted Self-Attention (CWSA). A fully connected Feed-Forward network is then followed as the other sub-layer. In both sub-layers, layer normalization\cite{layerNorm} and residual connection\cite{resnet} are employed to facilitate the robustness of GLOW Encoder.
\begin{small}
\begin{equation}
\begin{aligned}
    &{\rm CWSA} = {\rm Concat}({\rm WSA_1},...,{\rm WSA_n})W^o \\
    &{\rm CWSA_{out}}={\rm LayerNorm}({\rm CWSA}+{\rm X}) \\
    &{\rm GLOW Encoder}=\\
    &{\rm LayerNorm}({\rm CWSA_{out}}+{\rm FeedForward}({\rm CWSA_{out}}))
\end{aligned}
\end{equation}
\end{small}

\begin{algorithm}[t]
\caption{Encoding Algorithm}
\label{alg:A1}
\begin{algorithmic}[1]
\REQUIRE One query $q$, One document $d$, One vocabulary file $V_t$ for tokens, One idf vocabulary file $V_w$ for words, hyper-parameter $\alpha$ for BM25, hyper-parameter $\beta$ for BM25F.
\ENSURE The cosine similarity $s$ of this query document pair;\\

\COMMENT{/* Generating input features for all tokens, $W$ is word set, $w$ is word, $t$ is token, $S$ is segmentId set */}\\
\FOR{$w_i \in W$}
\FOR{$t_j \in w_i$}
\STATE $idf_j=V_w(w_i)$
\STATE $tf_j=tf(w_i)$
\STATE $tokenId_j=V_t(t_j)$
\STATE $segmentId_j=S(t_j)$
\ENDFOR
\ENDFOR\\

\COMMENT{/* Generating Encoded Vector for $d$ and $q$    */}
\STATE $v_q= {GLOW Encoder}^q(q, idf_q, tf_q, tokenId_q)$ \\
\STATE $v_d= {GLOW Encoder}^d(d, idf_d, tf_d, tokenId_d, fieldId_d)$ \\
\STATE $s=cosine(v_q, v_d)$

\end{algorithmic}
\end{algorithm}

We formulate the whole encoding logic in Algorithm~\ref{alg:A1}.

\begin{figure*}
\centering
\includegraphics[width=18cm]{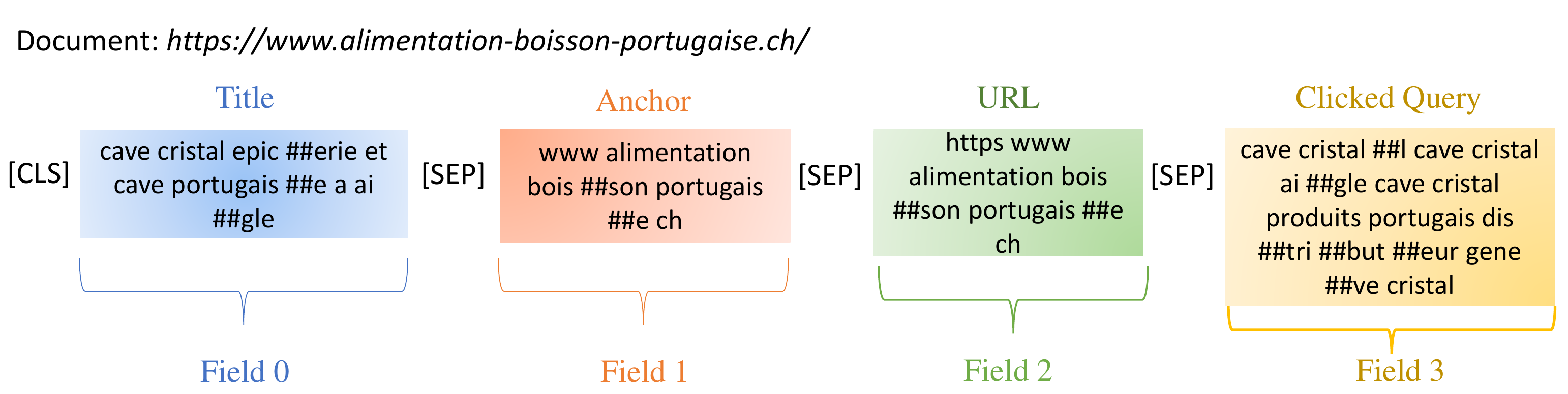}
\caption{Combined fields representation}
\label{fig:combinedfieldrepresentation}
\end{figure*}

\subsection{Global Weight Generation}
\label{weight_generation}
A key point of GLOW is to find an appropriate way to represent global weight. BM25 and its variants show the superiority in global weight representation for document ranking tasks against other alternatives. We leverage BM25 to generate the global weight scores for a query and BM25F\cite{Robertson1994OkapiAT} to compute the weight scores for a multi-fields document. 
BM25F is a modification of BM25 in which the document is considered to be composed from several fields with different degrees of importance in term of relevance saturation and length normalization. 
Both BM25 and BM25F depend on $tf$ and $idf$,
$tf$ means TermFrequency, it describes the number of occurrences of the word in the field. While $idf$ (InverseDocFrequency) is a measure of how much information the word provides, i.e., if it's common or rare across all documents. It is the logarithmically scaled inverse fraction of the documents that contain the word. For word $i$, $ idf_i = {\rm log}\frac{N-df_i+0.5}{df_i+0.5}$, where $N$ is a scalar \footnote{Here we set it by 100,000,000} indicting how many documents we are serving in system and $df$ is the number of documents where the word $i$ appears.
\paragraph{Inherent Query BM25} The calculation of classic BM25 is based on $tf$ in a document. Since in GLOW queries and documents are encoded separately,  here we compute an inherent query BM25 by computing $tf$ \textbf{within a query} instead. An inherent BM25 term weight for query word $i$ can be re-calculated as
\begin{equation}
 w_i^{BM25} = idf_i\frac{tf_i}{tf_i+k_1(1-b+b\frac{l_q}{avl_q})}
 \label{eq:queryBM25}
\end{equation}
where $tf_i$ is the term frequency of $word_i$ within query; $l_q$ is the query length; $avl_q$ is the query average length among all queries in the training set; $k_1$ is a free parameter usually chosen as 2 and 0<=$b$<=1 (commonly used is 0.75).
\paragraph{Inherent Document BM25F} In BM25F, instead of using $tf$ directly, empirically $atf$ (AdjustedTermFrequency) is widely adopted. It is proposed by adding several field-wise factors. For a $word_j$ in document field $c$, its $atf_j^c$ is defined in Eq. \ref{eq:atf}
\begin{equation}
 atf_j^c = \frac{fw_c\cdot{tf_j^c}}{1.0+fln_c\cdot{(\frac{fl_c}{avl_c}-1.0})}
 \label{eq:atf}
\end{equation}
where $fw_c$ is the weight of field $c$; $fln_c$ is the normalized field length for field $c$; $tf_j^c$ is the term frequency of $word_j$ within field $c$; $fl_c$ is the original field length of field $c$; $avl_c$ is the average length for field $c$.
\begin{equation}
 w_j^{BM25F} = idf_j\frac{atf_j}{k_1+atf_j}
 \label{eq:docBM25F}
\end{equation}
And its corresponding inherent BM25F score is computed in Eq. \ref{eq:docBM25F}, where the calculation of $idf_j$ is the same with $idf_i$.

\subsection{Whole Word Weight Sharing}
\label{wholeWordWeightShare}
Sub-word based approach has been proved efficient to alleviate out of vocabulary issue and limit vocabulary size. BERT uses WordPiece to produce tokens from original raw text. One shortcoming of this methodology is that we cannot directly apply the word-level prior knowledge. Moreover, in some NLP tasks, token-level weight is not enough to distinguish the importance of different words. The latest BERT model has proved that upgrading the Mask Language Model task to whole word level\footnote{https://github.com/google-research/bert} improves performance. In our task, the $W$ used for attention score is also based on whole word weights. That is, we first collect and calculate weights in whole word level, then give the same word weight to tokens corresponding to one word. By this way, one WordPiece token may has different weight representation if it occupies in different words. We also conduct experiment in Analysis section to compare the effect of token-level weight generation and word-level. The results suggest the word-level manner is superior than token-level.

\subsection{Combined Fields Representation}
\label{combinedFiledsRepresentation}
In ad-hoc retrieval tasks, there are always multiple sources of textual description (\textit{fields}) corresponding to one document. Lots of studies \cite{neuralrankingmodelmultiplefields,bm25ForWeightedFields} reveal that different fields contain complementary information. Thus, to obtain a more comprehensive understanding of document, when encoding the document into semantic vector space, we need to take multiple fields into consideration.

The well-known fields for a document in web search are title, header, keyword, body, and the URL itself etc. These fields are primitive from the website and can be fetched from HTML tags. Another kind of fields, like anchor, leverage the description from the brother website. Via this way, we can infer with useful information from other documents.
In addition to this, click signal is also with high quality and can be easily parsed from the search log. When a user clicked on the document \textit{d} with a query \textit{q}, we will add \textit{q} to the clicked query field of \textit{d}.

The special properties of these document fields make it difficult to unify them into one semantic space. One common approach \cite{neuralrankingmodelmultiplefields} is to separately encode the multiple fields respectively and learn a joint loss across these fields. Other alternatives unify all fields by simply concatenating all these fields with spaces.

As shown in Figure \ref{fig:combinedfieldrepresentation} we translate the segment definition of pre-next sentence in BERT to different fields in document by mapping multiple fields into different segments. Field embeddings are introduced to differentiate fields. Every field has a max length constrain, we set it to 20 tokens for \textit{anchor}, \textit{URL} and \textit{title} fields. For \textit{clicked query} fields, since a popular document may exist a large magnitude of click instances, we only pick the top 5 clicked queries for one document with a max length of 68 tokens. For all these fields, we pad them according to the need.
To obtain an unified document embedding, a [CLS] token is added at the beginning of the combined fields, and a [SEP] token is also inserted between each field to distinguish different fields.

\subsection{Objective Optimization}
\label{optimization}
We can achieve a sequence of semantic embeddings after GLOW Encoder. Inspired by \cite{devlin-etal-2019-bert}, using the embedding of [CLS] in the last layer as the matching features is already good enough. Nogueira et al.\cite{nogueira2019passage} also proved that in passage ranking task, adding more components upon Transformer does not help too much\cite{qiao2019understanding}. Therefore, GLOW uses the embedding of [CLS] as the semantic representation for the query and document, the matching score $s$ is measured by cosine similarity on the query and document vectors. 
\begin{equation}
    s = \rm{cos}(GLOW(query)_{cls}^{last},GLOW(document)_{cls}^{last})
\end{equation}
We adopt a binary cross entropy loss to optimize the model, which determines whether a query-document pair is relevant or not. We also tried pair-wise loss and found it had no extra improvement. Prior works \cite{qiao2019understanding,nogueira2019passage} also confirm on this.
\begin{equation}
    Loss = -y\rm{log}(\delta(w\cdot s+b))-(1-y)\rm{log}(1-\delta(w\cdot s+b))
\end{equation}
where $y$ is the label denoting if query-document is relevant, $\delta$ represents Sigmoid function. $w$ and $b$ are used to generate weighted cosine similarity to fit the Sigmoid function.

\section{Experiments}
To demonstrate whether and how our proposed GLOW framework can achieve performance improvements in the first-stage retrieval, we conduct comprehensive suite of experiments based on public dataset and real world search engine dataset including offline experiments, ablation study as well as case studies. Specifically, we break down the principal research questions into four individual ones to guide our experimentation.

\begin{itemize}
\item[\textbf{RQ1}] Can GLOW improve documents' quality comparing with the state-of-the-art retrieval models?
\item[\textbf{RQ2}] How does each component of GLOW contribute to its general performance?
\item[\textbf{RQ3}] Is GLOW better than BERT to capture the topical and thematic representation of the document?
\item[\textbf{RQ4}] Does GLOW increase model complexity and is it easier to converge to training data?
\end{itemize}

\subsection{Evaluation Datasets}
We evaluate the performance of GLOW on MS MARCO\footnote{The source code of our measurement on MS MARCO is available at \url{https://github.com/GLOW-deep/GLOW}} and Bing's large scale data points.
Table \ref{dataset_stat} illustrates the properties of the training data with these two datasets, it can be observed that the number of queries and documents in MS MARCO are within a pretty lower scale, only three hundred thousand queries and three million documents. It is hard to identify the deep matching model's quality for industrial web search engines with solely measurement on MS MARCO. Thus we expand the data scale by collecting more training data from Bing's search logs. In the Bing's internal data points, we scale up the query counts to 30 million and documents to 140 million. What's more, we also add clicked queries as the extra field in the internal set.


\begin{table}[h]
\centering
  \caption{Statistics of data sets used in our experiments. We use the development set of MS MARCO to evaluate model performance.}
  \label{dataset_stat}
  \small
  \begin{tabular}{l|l|l}
    \toprule
    Dataset  & MS MARCO & Intrinsic dataset\\  
     \midrule
     Train &367k queries, 36m q-d pairs & 30m queries, 310m q-d pairs\\
     \midrule
     Dev &5k queries, 3m documents & 14k queries, 70m documents\\
     \midrule
     Fields & url,title,body & anchor,title,url,clicked query\\
    \bottomrule
  \end{tabular}
\end{table}

\subsubsection{MS MARCO}
The document \textbf{full ranking} task in MS MARCO is similar with our scenario as the document contains multi fields with title and body. Based on our practice, the document URL can provide topical information, so we tokenize the raw URL into lexical representation as one more field.
In this task, training data contains 6 million query-document pairs with a simple binary positive/negative labeling while development set includes 5k queries and 3 million documents. For each query in evaluation set, we retrieve top 20 documents with the highest similarity scores for MRR calculation. 
\subsubsection{Instrinsic Bing's large scale dataset}
Similar with \cite{documentRankingwithDualWordEmbedding,dssm}
we sample 30 million real user queries and from real search engine logs. For the $training$ step, for each query, we treat its top 5 clicked documents as positives. By this way we obtain 150 million query-document pairs positive training examples. For the negatives, we use a mixture random sampling approach combining NCE negative sampling and Hard negative integration, which are detailed described below. These two methods consolidate the remaining 160 million negative query-document pairs.

\para{NCE negative sampling}
Directly random picking a negative case is too easy for the model to learn, which weakens the model's generalization. Instead, we use the noise-contrastive estimation (NCE) to pick competitive negatives\cite{nce,genericencoder}. It always picks negatives within current training batch with the same size of positives. 

\para{Hard negative integration}
The negatives from NCE sampling are all clicked documents, which only helps the model to learn entire non-related query-document pairs. To facilitate model with the capability to distinguish partial-related query-document pairs, we incorporate more difficult negatives by sampling 50 thousand queries from the search log and then sending these queries to the production system to retrieve 10 million partial-related query-document pairs as the hard negatives for these queries. These cases are added as companions of NCE negatives. 

The intrinsic evaluations are performed in a common used way\cite{intrinsicEvaluation} to evaluate the quality of semantic embedding representation. We pick 14k representative queries and 70 million documents candidates from the search logs as the development set.
Each query-document is human labelled with five standard categories in information retrieval(Perfect, Excellent, Good, Fair, Bad). 

\begin{table*}[h]
\centering
  \caption{Comparison of different deep matching models over MS MARCO dataset and Bing's internal data points. Full ranking task is picked for MS MARCO dataset and Dev set results are reported. XL-Net, BERT, BERT+DeepCT, BERT+Doc2query and GLOW are all trained with 3-layers Siamese architecture. The bold numbers indict the best result among other competitors and the superscript * denotes significant improvements over all the other models. }
  \label{intrinsic_evaluation}
  \begin{tabular}{l|ll|lll|lll}
    \toprule
    \multirow{2}{*}{Model} & \multicolumn{2}{c}{\textbf{MS MARCO (Dev)}}
    & \multicolumn{6}{|c}{\textbf{Bing Large Scale Query Set}} \\
    \cmidrule{2-9}
     &MRR@10 &MRR@20 &NDCG@1 &NDCG@3 &NDCG@10 &NCG@10 &NCG@20 &NCG@50 \\
    \midrule
    \texttt{TF-IDF} &0.1835 &0.1917 &0.1828 &0.2586 &0.3062 &0.3748 & 0.4561 &0.6274 \\
    \texttt{BM25} &0.2068 &0.2141 &0.2061 &0.2946 &0.3472 &0.4072 & 0.4889 &0.6574 \\
    \midrule
    \texttt{USE} &0.0627 &0.0648 &0.0860 &0.1003 &0.1171  &0.3548 & 0.4215 &0.6045\\
    \texttt{C-DSSM} &0.1461 &0.1506 &0.1900 & 0.2680 & 0.3113 &0.3845 & 0.4541 &0.6345 \\
    \midrule
    \texttt{XL-Net} &0.2597 &0.2659 &0.2528 &0.3515 &0.4017 &0.4217 &0.4947 &0.6541 \\
    \texttt{BERT Fine Tune} &0.2624 &0.2677 &0.3346 & 0.4567 &0.5154  &0.4255 &0.5054 & 0.6574 \\
    \midrule
    \texttt{BERT+DeepCT} &0.2677 &0.2725 &0.3364 &0.4598 &0.5275 &0.4425 &0.5147 &0.6748 \\
    \texttt{BERT+Doc2query} &0.2697 &0.2802 &0.3404 & 0.4621 &0.5298 &\textbf{0.4574} &0.5172 &0.6799\\
    \midrule
    \texttt{GLOW} &\textbf{0.2816} &\textbf{0.3104}\textsuperscript{*} &\textbf{0.3461} & \textbf{0.4772}\textsuperscript{*} &\textbf{0.5443}\textsuperscript{*} &0.4562 &\textbf{0.5284} &\textbf{0.7015}\textsuperscript{*}
    \\
    \bottomrule
  \end{tabular}
\end{table*}

\subsection{Experimental Setup}
\subsubsection{Evaluation metric}
To evaluate the effectiveness of the methods on MS MARCO, we use its official metric, Mean Reciprocal Rank(MRR) of the top-10 and 20 documents. MRR is a statistic measure for evaluating any process that produces a list of possible responses to a sample of queries, ordered by probability of correctness. 
For the Bing's large scale dataset, we adopt Normalized Discounted Cumulative Gain (NDCG)\cite{ndcg}, and report the results at position 1,3 and 10. Since we are focusing on the first stage retrieval in industrial web search, based on our experiences, Normalized Cumulative Gain(NCG) is another alternative to measure the matching documents' quality. Unlike NDCG, NCG considers more about the number of relevant documents got returned without caring the position because the second ranking stage is more responsible for the final ranking positions.
NCG is computed as
\begin{equation}
    \mathrm{NCG} = \frac{CG}{iCG}
\end{equation}
where Cumulative Gain(CG) is the sum of all the relevance scores in the ranking set.
\begin{equation}
\mathrm{Cumulative Gain(CG)} = \sum_{i=1}^{n} relevance_{i}   
\end{equation}
and ideal Cumulative Gain(iCG) is the sum of ideal document sets' CG.

\subsubsection{Baselines}
As mentioned in Section\ref{sec2}, our baselines contain classic information retrieval matching methods, typical deep learning models for sentence encoding, advanced pre-train language models and most recent vital benchmarks on document ranking.

\para{Classic retrieval methods.}
Given their deserved reputations in information retrieval history, we choose \texttt{TF-IDF} and \texttt{BM25} as representatives of the classic methods. \texttt{TF-IDF} is a numerical statistics that, by scoring the words in a text, indicates how important a word is in a document considering the corpus that document
belongs to. While \texttt{BM25} is a bag-of-words retrieval function that ranks a set of documents based on the query terms appearing in each document, regardless of their proximity within the document.

\para{Deep semantic models.}
Many primitive deep learning studies explored how to encode sentences into semantic embeddings. Among them, the Universal Sentence Encoder\cite{universalEncoder}(\texttt{USE}) and \texttt{C-DSSM}\cite{cdssm} are widely recognised to be more efficient and accurate. The former one leverages transfer learning to encode sentences
into embedding vectors while \texttt{C-DSSM} uses a convolutional-pooling structure over word
sequences to learn low-dimensional vector representations for search queries and Web documents.

\para{Pre-train language models.}
When stepping into language model pre-training boom, Transformer-based model has dominated the document ranking tasks. So we adopt \texttt{BERT} as one representative baseline of this group. \texttt{XL-Net}\cite{Yang2019XLNetGA}, a generalized autoregressive pretraining method, claims it achieves the state-of-the-art on document ranking task, we add it as the companion baseline with BERT.
We also refer prominent models in TREC deep learning track\footnote{https://microsoft.github.io/TREC-2020-Deep-Learning/}, such as DeepCT\cite{dai2019deepcpt} and  Doc2query\cite{doc2query} as the remaining baselines. The \texttt{BERT+DeepCT} baseline replaces the BERT component in DeepCT with context-independent word embeddings.
For a target document, Doc2query predicts a query, which can be treated as another new field to the document, hence the \texttt{BERT+Doc2query} baseline still takes BERT as base model with adding one more new field generated by Doc2query for all datasets.

\subsubsection{Training implementation}
To better accommodate industrial web search engine's demands, taking efficiency and scalability into consideration, also to make a fair comparison, we apply a \textbf{Siamese} framework with \textbf{3-layer} configuration for all aforementioned deep learning based benchmarks. All models are implemented using TensorFlow\footnote{\url{http://tensorflow.org/}}. We train GLOW with 8 Tesla V100 GPU, 32 GB memory for each. To best accelerate training, We implement a data parallel training pipeline base on horovod distribute training. What's more, Automatic Mixed Precision is enabled to train with half precision while maintaining the network accuracy. Finally, the training batch size is 500 query-document pairs. Adam optimizer\cite{kingma2014adam} is employed to train our model. The learning rate we used is 8e-5. We set the batch size to 300. Other hyber-parameters are the same with BERT. 
All of these competitor models are trained by following best practises suggested in previous works. They are evaluated strictly the same with GLOW by averaging the results collected from 5 times repeated training steps.

\begin{table}[h]
\centering
  \caption{Ablation evaluation comparison of GLOW variants with BERT and GLOW on Bing's large scale dataset. GLOW$_{idf}$ means using \textbf{i}nverse \textbf{d}ocument \textbf{f}requency as weight source. GLOW$_{wt}$ integrates the global \textbf{w}eights into original attention by a simple \textbf{t}rainable MLP layer.  GLOW$_{tw}$ indicates generating \textbf{t}oken level \textbf{w}eight for query and document. GLOW$_{us}$ uses an \textbf{u}nion \textbf{f}ield representation by setting all tokens' field embeddings as the same.}
  \label{ablation_evaluation}
  \small
   \resizebox{\linewidth}{!}
   {
      \begin{tabular}{l|lll|lll}
        \toprule
        Variant  & NDCG@1 &NDCG@3 & NDCG@10 &NCG@10 &NCG@20 &NCG@50\\  
         \midrule
         \texttt{BERT}&0.3346 &0.4567 &0.5154 &0.4255  &0.5054 & 0.6574   \\
         \midrule
        \texttt{GLOW$_{idf}$}&0.3349 &0.4587 &0.5149 &0.4397 & 0.5169 &0.6551   \\
         \midrule
        \texttt{GLOW$_{wt}$}&0.3330 &0.4625 &0.5134 &0.4269 & 0.5084 &0.6545   \\    
        \midrule
        \texttt{GLOW$_{tw}$}&0.3289 &0.4598 &0.5146 &0.4287 & 0.5011 & 0.6589   \\
        \midrule
        \texttt{GLOW$_{us}$}&0.3455 &0.4704 &0.5197 &0.4454 &0.5146 & 0.6898   \\
        \midrule
        \texttt{GLOW} &0.3461 &0.4772 &0.5243 &0.4502 &0.5204 &0.7015  \\
        \bottomrule
      \end{tabular}
  }
\end{table}

\subsection{Retrieval Quality Results (RQ1)}
\label{msmarco_exp}

To answer \textbf{RQ1}, we compare the retrival performance GLOW with aforementioned baselines on MS MARCO and Bing's large scale data points. The evaluation results are listed in Table \ref{intrinsic_evaluation}.

\para{MS MARCO results}
It can be observed from the left part of Table \ref{intrinsic_evaluation}, (1): \texttt{GLOW} shows the best performances on this dataset, it significantly beats \texttt{BERT} by a large margin of \textbf{7.3.\%} at MRR@10 and \textbf{15.9.\%} at MRR@20. (2): When increasing the retrieval count from 10 to 20, \texttt{GLOW} performs much better at MRR@20 than MRR@10. This trend indicates GLOW performing well when we want to fetch more ideal documents. (3): Even considering \texttt{BERT+DeepCT} and \texttt{BERT+Doc2query}, GLOW is also superior to them across all metrics. This result indicates that the manner \texttt{GLOW} performed with fusing global information into Transformer is effective for the first-stage retrieval.


\para{Intrinsic evaluation results}
Look at the right part of Table \ref{intrinsic_evaluation}, (1): Similar with results of MS MARCO, \texttt{GLOW} outperforms all benchmarks in metrics of NDCG@1,3,10 and NCG@10,50 and significantly gains in NDCG@3,10 and NCG@50 for this intrinsic dataset. (2) We see \texttt{BERT+Doc2query} accounts for the best result in NCG@10 but very close with \texttt{GLOW}, always better than \texttt{BERT+DeepCT} accross all metrics. This reflects that capturing documents' topical words might be more useful that assigning term weighting for queries in industrial large scale document retrieval.

The evaluations on MS MARCO and Bing's large scale dataset reveal that \texttt{GLOW} can not only address the real encoding problem for industrial search engine, but also extends to be a general solver for ordinary document matching task in IR community.

\begin{table*}[h]
    \caption{Top 5 words with highest attention scores from document between BERT and GLOW, they are selected from Url and Title fields. We fetch Url field by tokenization from raw web url. The green bold words in the table indicate the topical words while the red bold ones are off topic.}
    \label{attention_enhancement}
    \centering
    \scalebox{0.8}{
  \begin{tabular}{l|l|p{10cm}|p{4cm}|p{3.5cm}}
    \toprule
    Document & Field & Field Content& BERT top 5 words with highest attention & GLOW top 5 words with highest attention\\
	
	\midrule
	\multirow{2}{*}{Doc1} & Title & Top 30 Doctor insights on: Can   Low Sodium Levels Cause Seizures& 
	\colorbox{red!30}{\textbf{ insights}} low healthtap \colorbox{red!30}{\textbf{level}} cause& \colorbox{green!30}{\textbf{sodium}} healthetap \colorbox{green!30}{\textbf{seizures low}} insights \\& Url   & https www healthtap com topics   can low sodium levels cause seizures&&\\
	\midrule
	\multirow{2}{*}{Doc2} & Title & Police or Sheriff's Patrol Officer Salary& 
	payscale \colorbox{red!30}{\textbf{research}} police patro officer      & payscale \colorbox{green!30}{\textbf{sheriff}} 27s patrol \colorbox{green!30}{\textbf{ salary}}\\& Url   & https www payscale com research   us job police or sheriff 27s patrol officer salary  &&\\
	\midrule
	\multirow{2}{*}{Doc3} & Title & In pictures: Inside Hang Son Doong, the world's largest caves in Vietnam& pictures \colorbox{red!30}{\textbf{10914205 earth}} largest cases &caves \colorbox{green!30}{\textbf{hang son doong}} largest             \\& Url   & https www telegraph co uk news picturegalleries earth 10914205 in pictures inside hang son doong the worlds   largest caves in vietnam html &&\\
    \bottomrule
  \end{tabular}}
\end{table*}

\subsection{Ablation Study (RQ2)}
\label{ablationStudy}
To study \textbf{RQ2}, as aforementioned, three key components empower the embedding quality of GLOW. To further investigate the individual contribution of each part, we examine three GLOW variants and compare the NCG and NDCG results with BERT and GLOW full implementation on the Bing's internal dataset. Each variation disables a component while keeps others unchanged. The results reported in Table \ref{ablation_evaluation} indicate that all these three components are critical to  GLOW, missing anyone of them would degrade the performance of GLOW. Detailed analyses are described as follows.
\subsubsection{How to represent and integrate global weight?}
There are many alternatives to Eq.\ref{eq:weightedAttention} to serve as weight. One of common used ones in IR community is inverse document frequency (idf), it stands for the frequency across global documents for one particular word. However, idf only considers the global statistics without taking local context into consideration. Simply adopting idf as the global weight is easy to lead weight bias. To validate if BM25 is the best prior source for global weight estimation, we exploit a variant \texttt{GLOW$_{idf}$}, it use \textbf{i}nverse \textbf{d}ocument \textbf{f}requency as weight source.
\\GLOW combines the global weight with original attention with the multiplication. To validate if the multiplication is the best alternative to combine these two different weights, we design \texttt{GLOW$_{wt}$}, it integrates the global \textbf{w}eights into original attention by a simple \textbf{t}rainable MLP layer and outputs the final attention. 

Comparing the \texttt{BERT}, \texttt{GLOW$_{idf}$}, \texttt{GLOW$_{wt}$} and \texttt{GLOW} results in Table \ref{ablation_evaluation}, GLOW$_{idf}$ only performs slight better than BERT on NCG@10 and NCG@20, but nearly no change on NCG@50. Among all NDCG measurements, GLOW$_{idf}$ shows the flat trend. Besides,  GLOW$_{wt}$'s results are very close results with BERT on NCG and NDCG, largely drop comparing with \texttt{GLOW}. This indicates the representation and integration of global weight is crucial for \texttt{GLOW}, using improper weight representation or other integration manners could degrade the model performance.

\subsubsection{Whole word level weight $vs$ token level}
To verify the functionality of whole word weight sharing we proposed, we design \texttt{GLOW$_{tw}$}, it removes the whole word weight sharing by a straightforward \textbf{t}oken level \textbf{w}eight generation, which generates both query and document BM25 scores on WordPiece token level.

From Table \ref{ablation_evaluation}, there is nearly no improvement between \texttt{GLOW$_{tw}$} and \texttt{BERT} on all metrics, this proves computing BM25 on sub-word token level is not feasible. That is the reason why traditional search engine always use BM25 score on natural word level.

\subsubsection{Is field embedding a must?}
The token level representation of GLOW consists of 3 parts: token embeddings, position embeddings and field embeddings. Field embeddings are designed to differentiate fields for document. To validate if field embeddings are necessary. We design \texttt{GLOW$_{us}$}, which uses an \textbf{u}nion \textbf{f}ield representation by setting all tokens' field embeddings as the same. 

\texttt{GLOW$_{us}$} outperforms \texttt{BERT} and \texttt{GLOW$_{idf}$} and \texttt{GLOW$_{tw}$} on most metrics, but still have a slighter gap with GLOW Full, which signifies removing field embeddings will inevitably jeopardize the performance of \texttt{GLOW}.

\subsection{Case Study (RQ3)}
The design purpose of GLOW is to enhance the attention for those words having higher weights. We conduct some case studies to answer \textbf{RQ3}. Comparing BERT and GLOW, Table \ref{attention_enhancement} illustrates top 5 words which have the highest attention scores for each document. Here the attention scores are generated from the last layer's hidden state outputs, for each token, we firstly multiple all others by an inner product and  average the value, then sort all tokens' attention score. For those words belong to multiple tokens, we sum up the attention score to fetch those words' attention.

From Table \ref{attention_enhancement}, we can observe that GLOW is more effective to capture topical and semantic words from documents. Doc1 tries to share the answer to question 
"\textit{can low sodium levels cause seizures}", it is pretty clear here "\textit{low}", "\textit{sodium}" and "\textit{seizures}" are the topical words to these document. GLOW successfully rank these 3 words into Top5 while BERT only rank "\textit{low}" to \#2. In Doc2, "\textit{sheriff}" cannot rank into top 5 highest attention words in BERT, while it ranks second position in GLOW. This document has a strong preference on describing the officer salary information at $sheriff$ area, weakening this information makes the retrieval system harder to get this document back when people searches a query related with "\textit{sheriff patrol}". Similar phenomenon takes place at Doc3, BERT failed to emphasize "\textit{Hang Son Doong}”, which is vital to describe the topic of this document. But GLOW successfully enhance "\textit{Hang Son Doong}“'s attention. It is also explainable since "\textit{Hang Son Doong}” are all rare words, their global weights are pretty higher than other words, in addition, they repeat twice in this document so their BM25 value are higher than other words.

\subsection{Efficiency Analysis(RQ4)}
Then to answer \textbf{RQ4}, we evaluate the efficiency of GLOW from the points of view of model complexity and training cost. The training trend reveals GLOW is easier to converge while retaining the same training parameters with BERT.



\begin{figure}[h]
\centering
\begin{subfigure}{.25\textwidth}
  \centering
  \includegraphics[width=\textwidth]{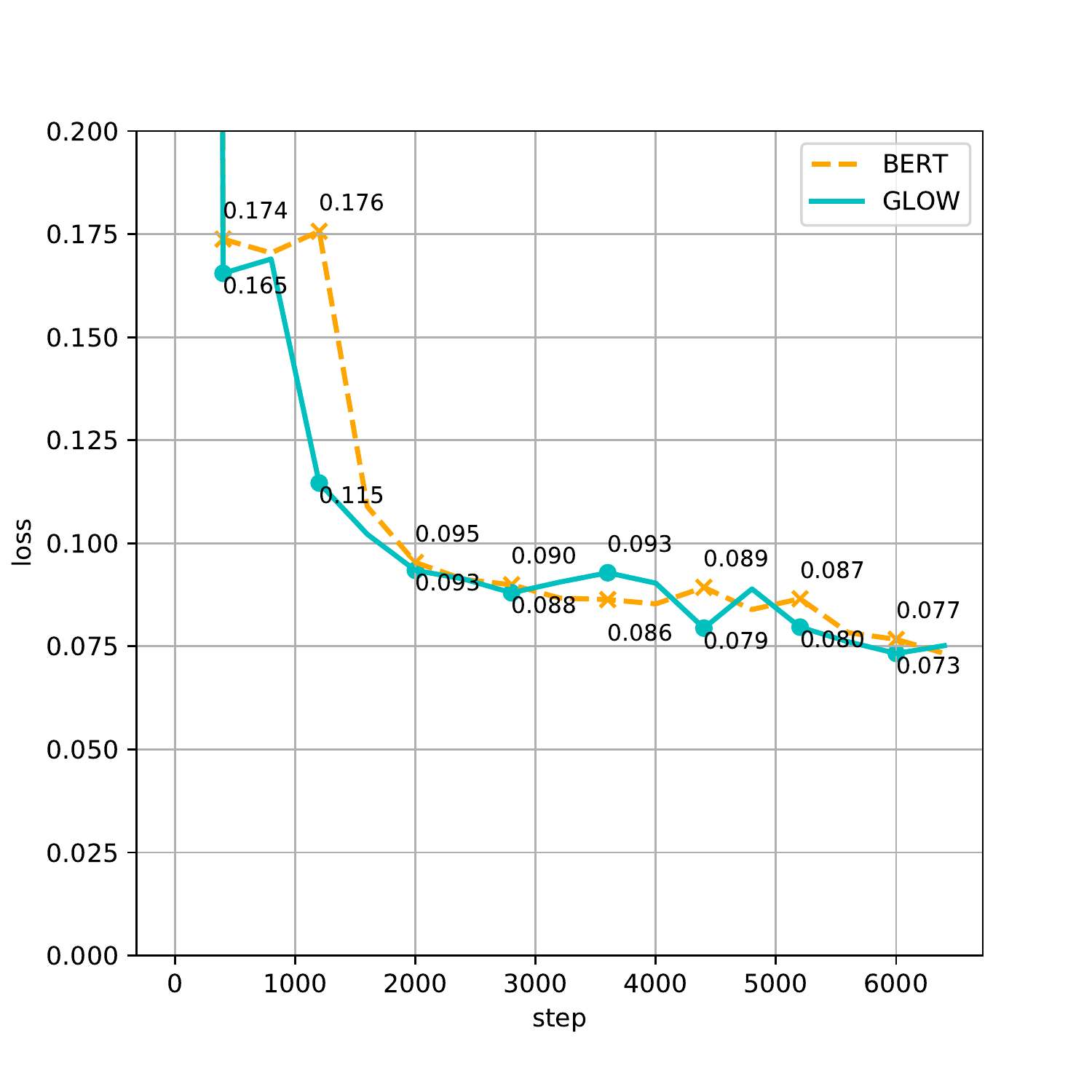}
  \captionof{figure}{Training loss}
  \label{fig:traing_loss}
\end{subfigure}%
\begin{subfigure}{.25\textwidth}
  \centering
  \includegraphics[width=\textwidth]{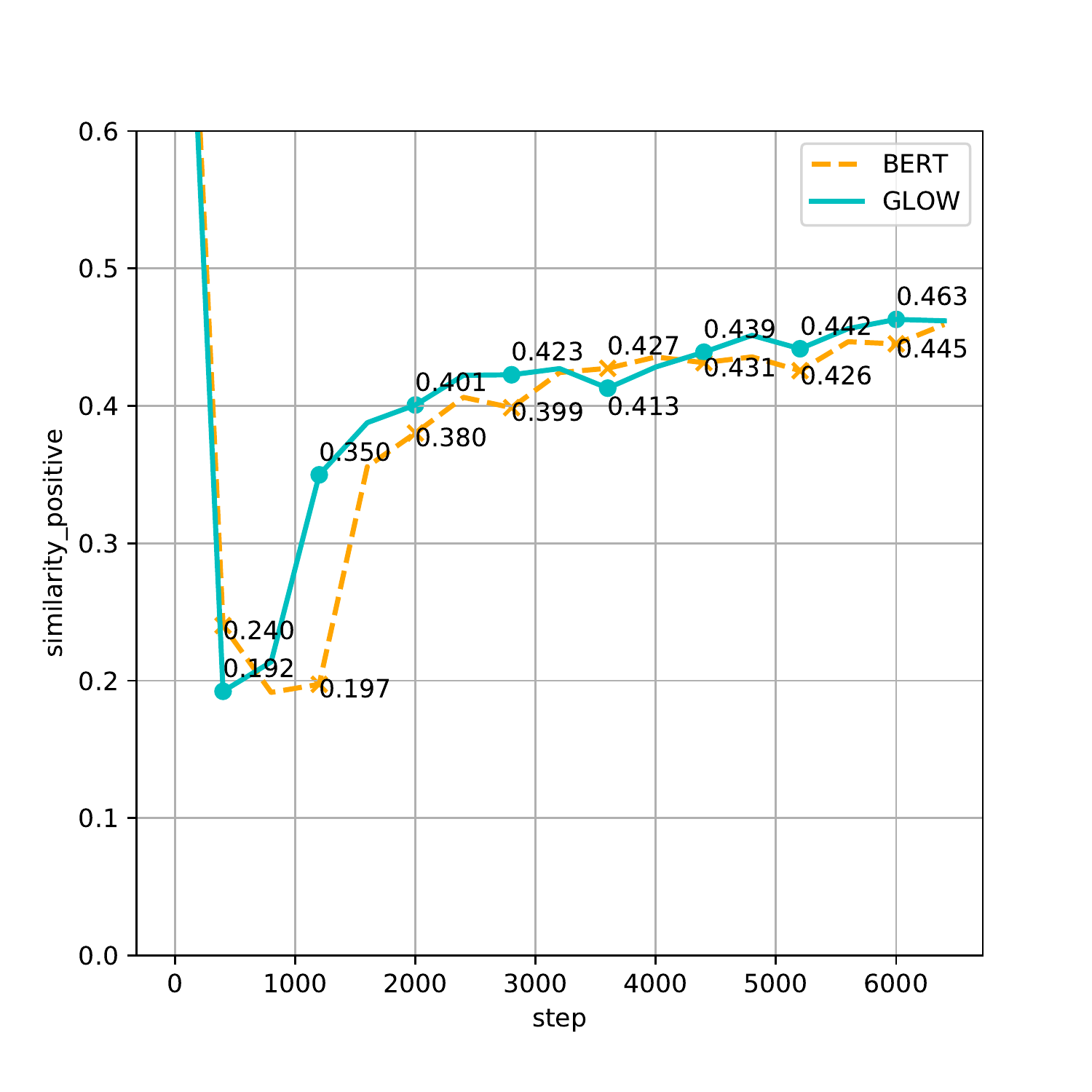}
  \captionof{figure}{Weighted cosine similarity}
  \label{fig:weighted_cosine_simi}
\end{subfigure}%
\caption{Efficiency comparison between BERT and GLOW. The yellow dotted lines represent BERT while green full lines indicate GLOW.}
\end{figure}

\para{Model complexity.}
One of the advantages for GLOW is that it does not involve any new trainable parameters. Thus it retains the same model complexity with BERT. The only extra work is to generate weight scores for all words which can be well prepared before model training and inference.

\para{Training efficiency.}
Researchers always expect deep models to be trained as fast as possible to reduce the training cost. As described in Sec \ref{optimization}, the value of weighted cosine similarity is important since it describes the discrimination between the query embeddings and document embeddings.
We plot the training loss and weighted cosine similarity trend curve of GLOW and BERT based on the training experience from MS MARCO.
According to Figure \ref{fig:traing_loss}, in the first 2k steps the training loss of GLOW decreases significantly faster than BERT. This indicates that GLOW is more easier to converge on training data, which could be a big time saving when we train document representation on a large scale data set. 
Practically, we always expect the weighted cosine similarity to be more differentiated to prevent the overlap across positives and negatives. Thus a large value of weighted cosine similarity is preferred. Figure \ref{fig:weighted_cosine_simi} shows that GLOW is superior in enlarging the weighted cosine similarity range rapidly.

\section{Conclusion}
We present GLOW, a general framework for matching phase in web search. It learns semantic representation for both queries and documents by integrating global weight into attention score calculation. By integrating the whole word weight sharing, it enhances whole word level attention interaction. Moreover, combined fields representation is proposed to fit GLOW into multi-fields document scenario. 
We conduct extensive experiments and rigorous analysis, demonstrating that GLOW outperforms other modern frameworks as the deep matching model. 





\bibliography{ref}
\bibliographystyle{plain}
\appendix

\end{document}